\documentclass{article}
\usepackage{spconf,amsmath,graphicx,amsfonts}

\title{BER Analysis of the box relaxation for BPSK Signal Recovery}
\name{Christos Thrampoulidis$^\star$, Ehsan Abbasi$^\star$, Weiyu Xu$^\dagger$, Babak Hassibi$^\star$}
\address{ 
$^\star$Department of Electrical Engeeniring, Caltech, Pasadena, USA \\
$^\dagger$Department of ECE, University of Iowa}
\usepackage[numbers]{natbib}

\usepackage{dsfont}

\usepackage{times}
\usepackage{hyperref}
\usepackage{url}
\usepackage{mathtools} 
\usepackage{cases}
\usepackage{subcaption}
\usepackage{amssymb,algorithm,cite,caption,cases,float,graphicx,url,color}
\usepackage{enumerate}
\usepackage{amsthm}

\definecolor{darkred}{RGB}{250,0,0}
\definecolor{darkgreen}{RGB}{0,150,0}
\definecolor{myblue}{RGB}{0,0,250}
\definecolor{darkblue}{RGB}{0,0,200}
\hypersetup{colorlinks=true, linkcolor=darkred, citecolor=myblue, urlcolor=darkblue}

\newboolean{showcomments}
\setboolean{showcomments}{true}
\newcommand{\ehsan}[1]{  \ifthenelse{\boolean{showcomments}}
{ \textcolor{blue}{(Bose says:  #1)}} {}  }
\newcommand{\chris}[1]{\ifthenelse{\boolean{showcomments}}
{ \textcolor{magenta}{(Chris says: #1)} } {} }
\newcommand{\babak}[1]{\ifthenelse{\boolean{showcomments}}
{ \textcolor{green}{(Babak says:  #1)}}{}}

\newcommand{\Pe}{P_e}
\newcommand{\BER}{BER~}

\usepackage{dsfont}
\newcommand{\ind}[1]{{\mathds{1}}_{\{#1\}}}
\newcommand{\gm}{\|\g\|}
\newcommand{\dB}{dB }
\newcommand{\BRO}{(BRO) }
\newcommand{\MFB}{(MFB) }
\newcommand{\PeMF}{P_e^{MFB}}

\newcommand{\Pro}{\mathbb{P}}
\newcommand{\psiubw}{\psi(\w,\ub)}
\newcommand{\Scc}{{\Sc^c}}

\theoremstyle{theorem}

\newtheorem{thm}{Theorem}[section]

\theoremstyle{remark}

\theoremstyle{definition}

\newcommand{\consist}{\citep[Thm.~2.7]{NF36} }
\newcommand{\uniform}{\citep[Cor..~II.1]{AG1982}}

\newcommand{\eps}{\epsilon}

\newcommand{\sign}{\mathrm{sign}}

\newcommand{\SNR}{\mathrm{SNR}}

\newcommand{\E}{\mathbb{E}}                    
\newcommand{\sigg}{\sigma^2}

\newcommand{\nn}{\notag}
\newcommand{\R}{\mathbb{R}}

\newcommand{\G}{\mathbf{G}}

\newcommand{\A}{\mathbf{A}}

\newcommand{\x}{\mathbf{x}}
\newcommand{\w}{\mathbf{w}}
\newcommand{\ub}{\mathbf{u}}
\newcommand{\g}{\mathbf{g}}
\newcommand{\vb}{\mathbf{v}}

\newcommand{\y}{\mathbf{y}}

\newcommand{\z}{\mathbf{z}}

\newcommand{\ab}{\mathbf{a}}

\newcommand{\h}{\mathbf{h}}

\newcommand{\Sc}{{\mathcal{S}}}

\newcommand{\Nn}{\mathcal{N}}

\newcommand{\beq}{\begin{equation}}
\newcommand{\eeq}{\end{equation}}
\newcommand{\bea}{\begin{align}}
\newcommand{\eea}{\end{align}}

\newcommand{\vp}{\vspace{4pt}}

\newcommand{\rP}{\xrightarrow{P}}

\begin{document}

%

\maketitle
\begin{abstract} 
We study the problem of recovering an $n$-dimensional vector of $\{\pm1\}^n$ (BPSK) signals from $m$ noise corrupted measurements $\y=\A\x_0+\z$. In particular, we consider the box relaxation method which relaxes the discrete set $\{\pm1\}^n$ to the convex set $[-1,1]^n$ to obtain a convex optimization algorithm followed by hard thresholding. 
When the noise $\z$ and measurement matrix $\A$ have iid standard normal entries, we obtain an exact expression for the bit-wise probability of error $\Pe$ in the limit of $n$ and $m$ growing and $\frac{m}{n}$ fixed. At high SNR our result shows that the $\Pe$ of box relaxation is within $3$\dB of the matched filter bound \MFB for square systems, and that it approaches \MFB as $m $ grows large compared to $n$.  Our results also indicates that as $m,n\rightarrow\infty$, for any fixed set of size $k$, the error events of the corresponding $k$ bits in the box relaxation method are independent.

\end{abstract}
 


\section{Introduction}
The problem of recovering an unknown BPSK vector from a set of noise corrupted linearly related measurements arises in numerous applications, such as Massive MIMO \citep{ngo2013energy,wen2014message,narasimhan2014channel,MIMO_maleki}. As a result, a large host of exact and heuristic optimization algorithms have been proposed. Exact algorithms, such as sphere decoding and its variants, become computationally prohibitive as the problem dimension grows. Heuristic algorithms such as zero-forcing, MMSE, decision-feedback, etc., \citep{grotschel2012geometric,foschini1996layered,hassibi2005sphere} have inferior performances that are often difficult to precisely characterize.

One popular heuristic is the so called "Box Relaxation" which replaces the discrete set $\{\pm1\}^n$ with the convex set $[-1,1]^n$ \citep{boyd2009convex}. This allows one to recover the signal via convex optimization followed by hard thresholding. Despite its popularity, very little is known about the performance of this method. In this paper, we exactly characterize its bit-wise error probability in the regime of large dimensions and under Gaussian assumptions.

The remainder of the paper is organized as follows. Some background and the formal problem definition are given in the rest of this section. The main result and a detailed discussion follows in Section \ref{sec:result}. An outline of the proof is the subject of Section \ref{sec:proof}. Finally, the paper concludes in Section \ref{sec:conc}.

\subsection{Setup}

Our goal is to recover an $n$-dimensional BPSK vector $\x_0\in\{\pm1\}^n$ from the noisy multiple-input multiple output (MIMO) relation 
$\y=\A\x_0+\z\in\R^m,$
where $\A\in\R^{m\times n}$ is the MIMO channel matrix (assumed to be known) and $\z\in\R^m$ is the noise vector. We assume that $\A$ has entries iid $\Nn(0,1/n)$ and $\z$ has entries iid $\Nn(0,\sigma^2)$. The normalization is such that the reciprocal of the noise variance $\sigma^2$ is equal to the Signal-to-Noise Ratio, i.e.
$
\SNR = 1/{\sigma^2}.
$

\vp
\noindent\textbf{The Maximum-Likelihood (ML) decoder}. The ML decoder which maximizes the probability of error (assuming the $\x_{0,i}$ are equally likely) is given by 
$
\min_{\x\in\{\pm1\}^n}\|\y-\A\x\|_2.
$
 Solving the above, is often computationally intractable, especially when $n$ is large, and therefore a variety of heuristics have been proposed (zero-forcing, mmse, decision-feedback, etc.) \citep{verdu1998multiuser}.
 
\vp
\noindent\textbf{Box Relaxation Optimization}.
The heuristic we shall use, we  refer to it as Box Relaxation Optimization (BRO). It consists of two steps. The first one involves solving a convex relaxation of the ML algorithm, where $\x\in\{\pm1\}^n$ is relaxed to $\x\in[-1,1]^n$. The output of the optimization is  hard-thresholded in the second step to produce the final binary estimate. Formally, the algorithm outputs an estimate 
$\x^*$
 of $\x_0$ given as
 \vspace{-5pt}
\begin{subequations}\label{eq:algo}
\begin{align}
\hat\x = \arg\min_{-1\leq\x_i\leq 1}\|\y-\A\x\|_2,\label{eq:LASSO}\\
\x^* = \sign(\hat\x),
\end{align}
\end{subequations}
where the $\sign$ function returns the sign of its input and acts element-wise on input vectors.

\vp
\noindent\textbf{Bit error probability}.
 We evaluate the performance of the detection algorithm by the bit error probability $\Pe$,  defined as the expectation of the Bit Error Rate $\BER$. Formally,
 \vspace{-5pt}
\begin{subequations}
\begin{align}
\BER &:=  \frac{1}{n}\sum_{i=1}^n \ind{\x^*_i\neq \x_{0,i}}, \label{eq:BER}\\
\Pe := \E\left[\BER \right] &= \frac{1}{n}\sum_{i=1}^n \Pr\left(\x^*_i\neq \x_{0,i}\right).\label{eq:Pe}
\end{align}
\end{subequations}

\vspace{-10pt}

\section{Main Result}\label{sec:result}
Our main result analyzes the $\Pe$ of the \BRO in \eqref{eq:algo}. We assume a large-system limit where  $m,n\rightarrow\infty$ at a proportional rate $\delta$. The $\SNR$ is assumed constant; in particular, it does not scale with $n$. Let $Q(\cdot)$ denote the  Q-function associated with the standard normal density $p(h)=\frac{1}{\sqrt{2\pi}}\mathrm{e}^{-h^2/2}$.

\begin{thm}[$\Pe$ of the \BRO]\label{thm:main}
Let $\Pe$ denote the bit error probability of the detection scheme in \eqref{eq:algo} for some fixed but unknown BPSK signal $\x_0\in\{\pm1\}^n$. For constant $\SNR$ and $\frac{m}{n}\rightarrow\delta\in(\frac{1}{2},\infty)$, it holds:
$$
\lim_{n\rightarrow\infty}\Pe = Q(1/\tau_*),
$$
where $\tau_*$ is the unique solution to 
\begin{align}\label{eq:DO_thm}
\min_{\tau>0}~\frac{\tau}{2}\left(\delta-\frac{1}{2}\right)+\frac{1/\SNR}{2\tau}-\frac{\tau}{2}\int_{\frac{2}{\tau}}^{\infty}\left(h+\frac{2}{\tau}\right)^2p(h)\mathrm{d}h.
\end{align}
\end{thm}
\vspace{-10pt}
Theorem \ref{thm:main} derives a \emph{precise} formula for the bit error probability of the (BRO). The formula involves solving a \emph{convex} and deterministic minimization problem in \eqref{eq:DO_thm}.
We outline the proof in Section \ref{sec:proof}. First, a few remarks are in place.

\vspace{-10pt}
\subsection{Remarks}
\vspace{-10pt}

\vp\noindent\textbf{Computing $\tau_*$}. It can be shown that the objective function of \eqref{eq:DO_thm} is strictly convex when $\delta>\frac{1}{2}$.
When $\delta<\frac{1}{2}$, it is well known that even the noiseless box relaxation fails \citep{Cha}. (In fact, $\delta=\frac{1}{2}$ is the recovery threshold for this convex relaxation.)
 Thus, \eqref{eq:DO_thm} has a unique solution $\tau_*$. Observe that the problem parameters $\delta$ and $\SNR$ appear explicitly in \eqref{eq:DO_thm}; naturally then $\tau_*$ is indeed a function of those. The minimization in \eqref{eq:DO_thm} can be efficiently solved numerically. In addition, owing to the strict convexity of the objective function, $\tau_*$ can be equivalently expressed as the unique solution to the corresponding first order optimality conditions. 

\begin{figure}[t]
    \centering
    \includegraphics[width=0.5\textwidth]{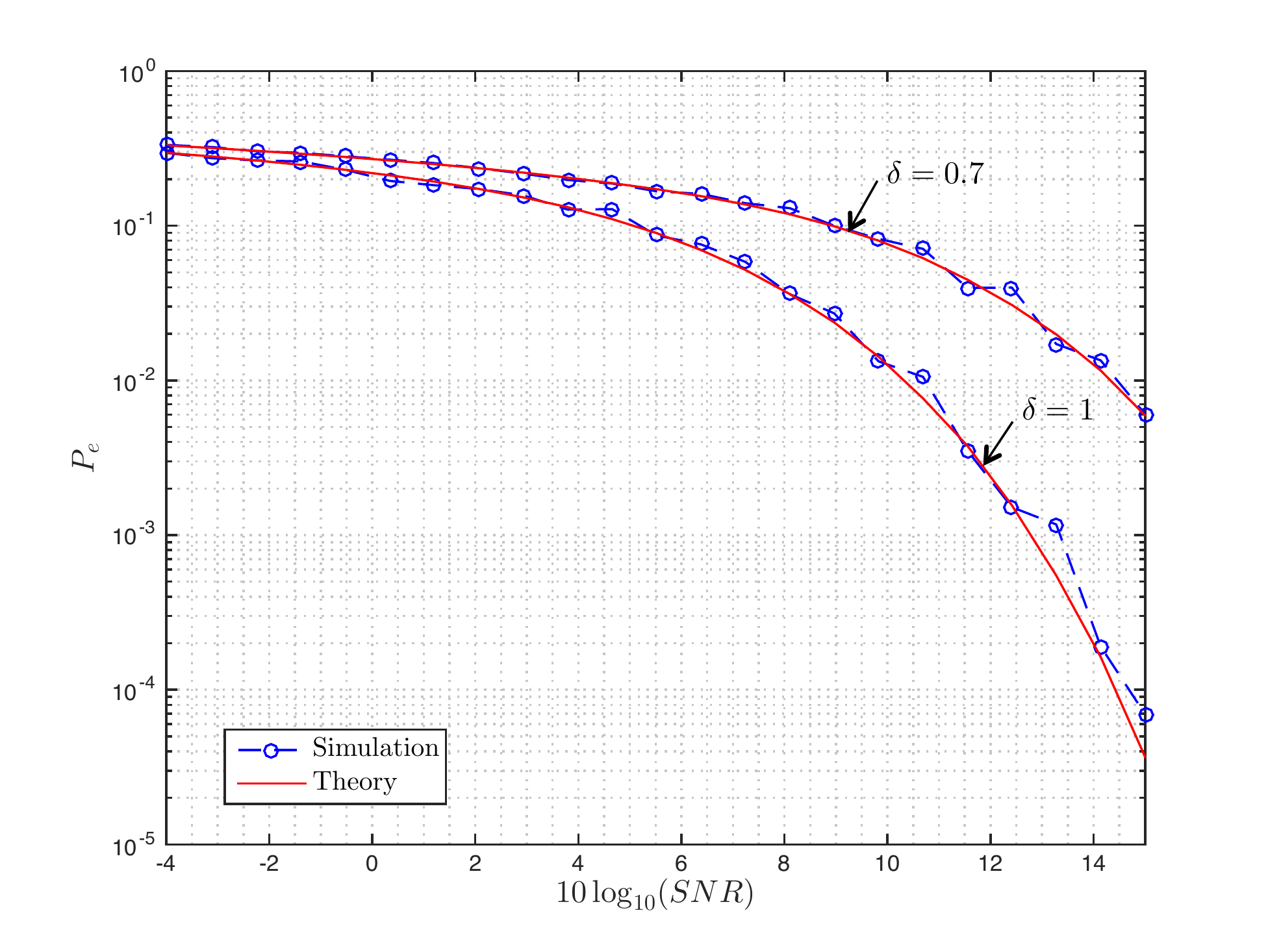}
    \caption{\footnotesize{BER Performance of the Boxed Relaxation: $\Pe$ as a function of $\SNR$ for different values of the ration $\delta=\lceil m/n \rceil$. The theoretical prediction follows from Theorem \ref{thm:main}. For the simulations, we used $n=512$. The data are averages over 20 independent realizations of the channel matrix and of the noise vector for each value of the $\SNR$. }}
    \label{fig:sim}
\end{figure}

\vp\noindent\textbf{Numerical illustration}. Figure \ref{fig:sim} illustrates the accuracy of the prediction of Theorem \ref{thm:main}. Note that although the theorem requires $n\rightarrow\infty$, the prediction is already accurate for $n$ ranging on a few hundreds.

\vp\noindent\textbf{Analysis of convex algorithms}. We are able to predict the $\Pe$ of the detection algorithm in \eqref{eq:algo} by analyzing the performance of the convex algorithm in \eqref{eq:LASSO}. These type of convex algorithms, which minimize a least-squares function of the residual $\y-\A\x$ subject to a (typically non-smooth) constraint on $\x$, are often referred to in the (statistics and signal-processing) literature as LASSO-type algorithms. When the performance of such algorithms is measured in terms of the square-error $\|\hat\x-\x_0\|_2^2$, the recent line of works \citep{StoLASSO,OTH13,ICASSP15,COLT} has led to precise results and a clear understanding of its asymptotic behavior. 
The analysis of these works builds upon the Convex Gaussian Min-max Theorem (CGMT) \citep[Thm.~1]{COLT}, which is an extension of a classical Guassian comparison inequality due to Gordon \citep{GorThm}. Of interest to us is not calculating the squared-error of \eqref{eq:LASSO} but rather the $\Pe$.  Thus, we manage to extend the precise analysis of the LASSO-type algorithm beyond the squared-error. To prove our result we require a slight generalization of the CGMT as it appears in \citep{COLT}.

%

\begin{figure}[t]
    \centering
    \includegraphics[width=0.5\textwidth]{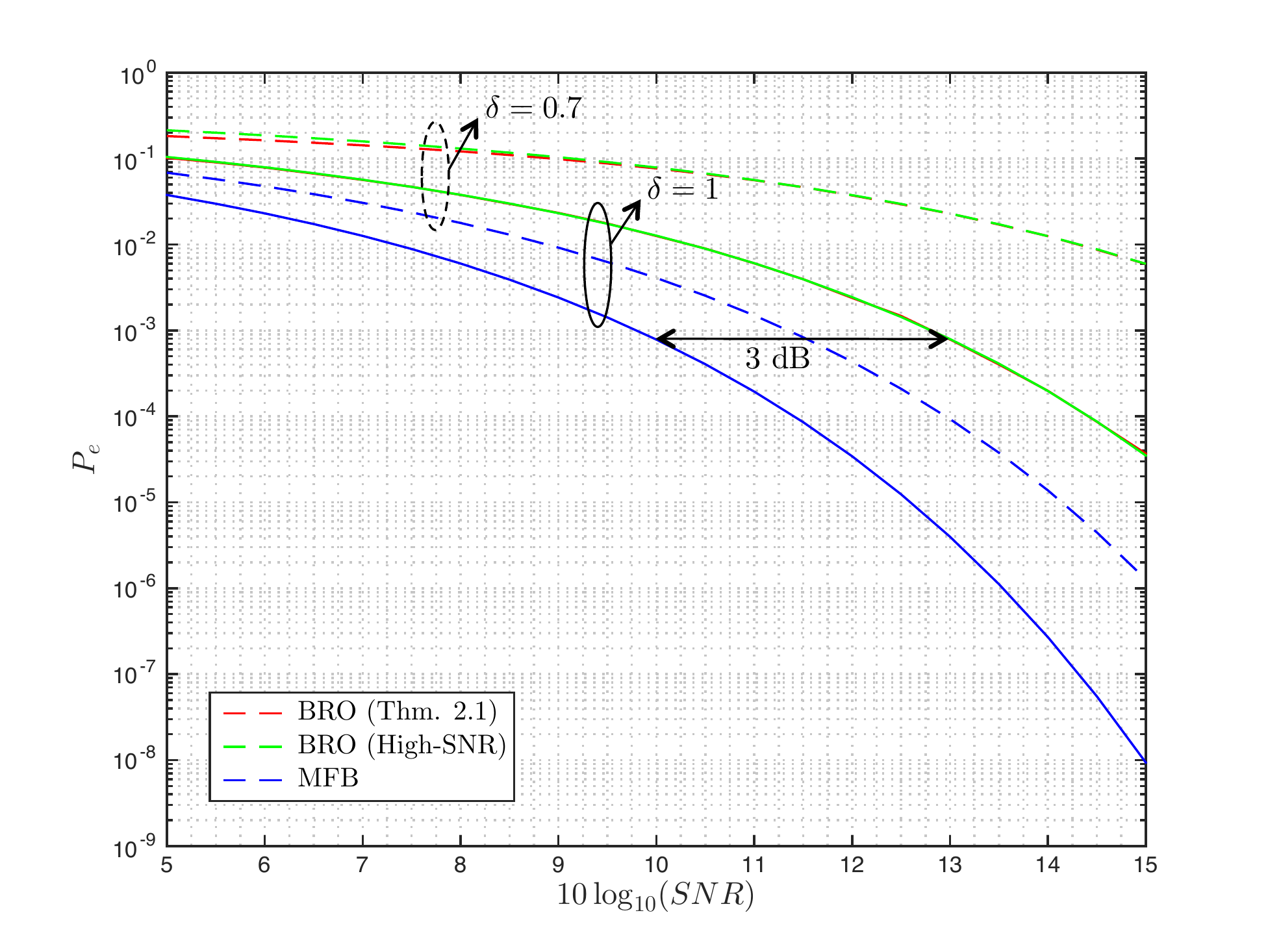}
    \caption{ \footnotesize{Bit error probability of the Box Relaxation Optimization (BRO) in \eqref{eq:algo} in comparison to the Matched Filter Bound (MFB) for $\delta=0.7$ (dashed lines) and $\delta=1$ (solid lines). The red curves follow the formula of Thm.~\ref{thm:main}, the green ones correspond to \eqref{eq:BRO_high}, and, $\PeMF$ of \eqref{eq:MFB_high} is  in blue. 
    }
    }
    \label{fig:compare}
\end{figure}

\vp\noindent\textbf{$\Pe$ at high-SNR}. 
It can be shown that when $\SNR\gg1$, then $\tau_*=1/\sqrt{(\delta-1/2)SNR}$. This can be intuitively understood as follows: at high-SNR, we expect $\tau_*$ to be going to zero (correspondingly $\Pe$ to be small). When this is the case, the last term in \eqref{eq:DO_thm} is negligible; then, $\tau_*$ is the solution to $\min_{\tau>0}\frac{\tau}{2}\left(\delta-\frac{1}{2}\right)+\frac{1/\SNR}{2\tau}$ which gives the derided result.
Hence, for $\SNR\gg1$,
\vspace{-5pt}
\begin{align}\label{eq:BRO_high}
\lim_{n\rightarrow\infty}\Pe \approx Q(\sqrt{\left(\delta-1/2\right)\cdot\SNR}).
\end{align}
In Figure \ref{fig:compare} we have plotted this high-SNR expression for the $\log_{10}(\Pe)$ vs its exact value as predicted by Theorem \ref{thm:main}. It is interesting to observe that the former is actually a very good approximation to the latter even for small practical values of SNR. The range of SNR values for which the approximation is valid becomes larger with increasing $\delta$. Heuristically, for $\delta>0.7$ the expression in \eqref{eq:BRO_high} is a good proxy for the true probability of error at practical SNR values.

\vp\noindent\textbf{Comparison to the matched filter bound}. Theorem \ref{thm:main} gives us a handle on the $\Pe$ of \BRO in \eqref{eq:algo} and therefore allows to evaluate its practical performance. Here, we compare the performance to an idealistic case, where all $n-1$, but $1$, bits of $\x_0$ are known to us. As is customary in the field, we refer to the bit error probability of this case as the \emph{matched filter bound} \MFB and denote it by $\PeMF$. The \MFB corresponds to the probability of error in detecting (say) $\x_{0,n}\in\{\pm1\}$ from:
$
\tilde\y = \x_{0,n}\ab_n + \z,
$
where  $\tilde\y=\y-\sum_{i=1}^{n-1}\x_{0,i}\ab_i$ is assumed known, and, $\ab_i$ denotes the $i^{th}$ column of $\A$.
The ML estimate is just the sign of the projection of the vector $\tilde{\y}$ to the direction of $\ab_n$. Without loss of generality assume that $\x_{0,n}=1$. Then, the output of the matched filter becomes $\sign(\tilde{X})$, where
$
\tilde{X} = \|\ab_n\|^2+\sigma^2\nu,
$
where $\nu\sim\Nn(0,1)$. When $n\rightarrow\infty$, $\|\ab_n\|^2\rP \delta$\footnote{We use $\rP$ to denote convergence in probability with $n\rightarrow\infty$.} . Hence, with probability one,
\begin{align}\label{eq:MFB_high}
\lim_{n\rightarrow\infty}\PeMF = \lim_{n\rightarrow\infty}\Pro(\tilde X<0) = Q(\sqrt{\delta\cdot\SNR}).
\end{align}
A direct comparison of \eqref{eq:MFB_high} to \eqref{eq:BRO_high} shows that
at high-SNR, the performance of the \BRO is $10\log_{10}\frac{\delta}{\delta-1/2}$\dB off that of the \MFB. In particular, in the square case ($\delta=1$), where the number of receive and transmit antennas are the same, the \BRO is 3\dB off the \MFB. When the number of receive antennas is much larger, i.e. when $\delta\rightarrow\infty$, then the performance of the \BRO approaches the \MFB.

%


\vspace{-8pt}
\section{Proof Outline}\label{sec:proof}
\vspace{-2pt}
For simplicity, we write $\|\cdot\|$ for the $\ell_2$-norm.

\vp\noindent\textbf{The error vector}. 
It is convenient to re-write  \eqref{eq:LASSO} by changing the variable to the error vector $\w:=\x-\x_0$:
\vspace{-5pt}
\begin{align}\label{eq:LASSOw}
\hat\w := \min_{-2\leq\w_i\leq 0} \|\z-\A\w\|.
\end{align}
\vspace{-2pt}
Without loss of generality we assume for the analysis that $\x_0=\mathbf{1}_n=(1,1,\ldots,1)$. Then, we can write \eqref{eq:BER} in terms of the error vector $\w$ as:
$\BER=\frac{1}{n}\sum_{i=1}^n\ind{\hat\w_i\leq -1}.$

\vp\noindent{\textbf{The CGMT}}.
The fundamental tool behind our analysis is the Convex Gaussian Min-max Theorem (CGMT) \citep{COLT}; CGMT 
builds upon a classical result due to Gordon \citep{GorThm}\footnote{Gordon's original result is classically used to establish non-asymptotic probabilistic lower bounds on the minimum singular value of Gaussian matrices (e.g. \citep{Ver}), and has a number of other applications in high-dimensional convex geometry (e.g. \citep{GorLem,ledoux}).}. It  associates with a primary optimization (PO) problem a simplified auxiliary optimization (AO) problem from which we can tightly infer properties of the original (PO), such as the
optimal cost, the optimal solution, etc.. 
The idea of combining the GMT with convexity is attributed to Stojnic \citep{StoLASSO}. Thrampoulidis, Oymak and Hassibi built and significantly extended on this idea arriving at the CGMT as it appears in \citep[Thm.~3]{COLT}
For the ease of reference we repeat a statement of the CGMT here; in this generality the theorem appears in \citep{Master}. 
Consider the following two min-max  problems
\begin{subequations}\label{eq:POAO}
\begin{align}
\label{eq:PO_gen}
\Phi(\G)&:= \min_{\w\in\Sc_\w}~\max_{\ub\in\Sc_\ub}~ \ub^T\G\w + \psiubw,\\
\label{eq:AO_gen}
\phi(\g,\h)&:= \min_{\w\in\Sc_\w}~\max_{\ub\in\Sc_\ub}~ \|\w\|\g^T\ub - \|\ub\|\h^T\w + \psiubw,
\end{align}
\end{subequations}
where $\G\in\R^{m\times n}, \g\in\R^m, \h\in\R^n$, $\Sc_\w\subset\R^n,\Sc_\ub\subset\R^m$ and $\psi:\R^n\times\R^m\rightarrow\R$. Denote $\w_\Phi:=\w_\Phi(\G)$ and $\w_\phi:=\w_\phi(\g,\h)$ any optimal minimizers in \eqref{eq:PO_gen} and \eqref{eq:AO_gen}, respectively. 
%
\begin{thm}[CGMT]\label{thm:CGMT}
In \eqref{eq:POAO}, let $\Sc_\w,\Sc_\ub$ be convex and compact sets, $\psi$ be continuous and convex-concave on $\Sc_\w\times\Sc_\ub$, and, $\G,\g$ and $\h$ all have entries iid standard normal. 
Let $\Sc$ be an arbitrary open subset of $\Sc_\w$ and $\Scc=\Sc_\w/\Sc$. Denote $\phi_\Scc(\g,\h)$ the optimal cost of the optimization in \eqref{eq:AO_gen}, when the minimization over $\w$ is now constrained over $\w\in\Scc$. 
 If in the limit of $n\rightarrow\infty$  both $\phi(\g,\h)$ and $\phi_\Scc(\g,\h)$ converge in probability, and, 
 $\lim_{n\rightarrow\infty}\Pr(\w_\phi \in \Sc)=1$, then, it also holds   $\lim_{n\rightarrow\infty}\Pr(\w_\Phi \in \Sc)=1$.

%
%
\end{thm}

\vspace{-5pt}
\vp\noindent\textbf{Identifying the (PO) and the (AO)}. 
Using the CGMT for the analysis of the  $\Pe$, requires as a first step  expressing the optimization in \eqref{eq:LASSO} in the form of a (PO) as it appears in \eqref{eq:PO_gen}. It is easy to see that \eqref{eq:LASSOw} is equivalent to 
\vspace{-5pt}
\begin{align}
\min_{-2\leq\w_i\leq 0}\max_{\|\ub\|\leq 1} \ub^T\A\w-\ub^T\z.
\end{align}
\vspace{-2pt}
Observe that the constraint sets above are both convex and compact; also, the objective function is convex in $\w$ and concave in $\ub$. Hence, according to the CGMT we can perform the analysis of the $\BER$ for the corresponding (AO) problem instead, which becomes (note the normalization to account for the variance of the entries of $\A$)
\vspace{-5pt}
\begin{align}\label{eq:AOw}
\frac{1}{\sqrt{n}}\min_{-2\leq\w_i\leq 0}\max_{\|\ub\|\leq 1} (\|\w\|\g-\sqrt{n}\z)^T\ub-\|\ub\|\h^T\w.
\end{align}
We refer to the optimization in \eqref{eq:AOw} as the (AO) problem. 

\vp\noindent\textbf{Computing the $\BER$ via the (AO)}.
Call $\tilde\w$ the optimal solution of the (AO).
Fix any $\eps>0$ and consider the set 
\begin{align}\label{eq:Sc}
\Sc=\{\vb : \Big|\frac{1}{n}\sum_{i=1}^n\ind{\vb_i\leq -1} - Q(1/\tau_*)\Big|<\eps \},
\end{align}
where $\tau_*$ is defined in the statement of Theorem \ref{thm:main}.
 We will apply Theorem \ref{thm:CGMT} for the above set $\Sc$. In particular, we  show that  (i) the (AO) in \eqref{eq:AOw} converges in probability (after proper normalization with $n$), and, (ii) $\tilde\w\in\Sc$ with probability one. These will suffice to conclude that $\hat\w\in\Sc$ with probability one, which would complete the proof of Theorem \ref{thm:main}.

\vp\noindent\textbf{Simplifying the (AO)}.
  We begin by simplifying the (AO) problem as it appears in \eqref{eq:AOw}. First, since both $\g$ and $\z$ have entries iid Gaussian, then, $\|\w\|\g-\sqrt{n}\z$ has entries iid $\Nn(0,\sqrt{\|\w\|^2+n\sigma^2})$. Hence, for our purposes and using some abuse of notation so that $\g$ continues to denote a vector with iid standard normal entries, the first term in \eqref{eq:AOw} can be treated as $\sqrt{\|\w\|^2+n\sigma^2}\g^T\ub$, instead. As a next step,
fix the norm of $\ub$ to say $\|\ub\|=\beta$. Optimizing over its direction is now straightforward, and gives
$\min_{-2\leq\w_i\leq 0}~\max_{0\leq \beta\leq 1} \frac{\beta}{\sqrt{n}}\left(\sqrt{\|\w\|^2+n\sigma^2}\|\g\|-\h^T\w\right).$
In fact, it is easy to now optimize over $\beta$ as well; its optimal value is $1$ if the term in the parenthesis is non-negative, and, is 0 otherwise. With this, the (AO) simplifies to the following:
\begin{align}
\big(\min_{-2\leq\w_i\leq 0}\sqrt{\frac{\|\w\|^2}{n}+\sigma^2}\|\g\|-\frac{1}{\sqrt{n}}\h^T\w\big)_+,\nn
\end{align}
where we defined $(\chi)_+:=\max\{\chi,0\}$.
To facilitate the optimization over $\w$, we express the term in the square-root in a variational form, using $\sqrt{\chi}=\min_{\tau>0}\frac{\tau}{2} + \frac{\chi}{2\tau}$. With this trick, the minimization over the entries of $\w$ becomes separable:
\begin{align}
\min_{\substack{\tau\geq 0}}\frac{\tau\|\g\|}{2} + \frac{\sigg\|\g\|}{2\tau}+ \sum_{i=1}^n \min_{-2\leq\w_i\leq 0} \frac{\gm}{2\tau n}\w_i^2-\frac{\h_i}{\sqrt{n}}\w_i.\nn
\end{align}
Then, 
the optimal $\tilde\w_i$ satisfies
\vspace{-5pt}
\begin{align}\label{eq:w_opt}
\tilde\w_i = \begin{cases}
0&,\text{if } \h_i\geq 0,\\
\frac{\tau\sqrt{n}}{\gm}\h_i&,\text{if } -\frac{2\gm}{\tau\sqrt{n}}\leq \h_i < 0, \\
-2&,\text{if }  \h_i< -\frac{2\gm}{\tau\sqrt{n}}.
\end{cases}
\end{align}
where, $\tau$ is the solution to the following:
%
\vspace{-5pt}
\begin{align}\label{eq:AOw3}
\left(\min_{\tau>0} \frac{\tau\|\g\|}{2} + \frac{\sigg\gm}{2\tau} + \frac{1}{\sqrt{n}}\sum_{i=1}^n \upsilon(\tau;\h_i,\gm) \right)_+, 
\end{align}
\vspace{-5pt}
\begin{align}
\upsilon(\tau;\h_i,\gm) :=\begin{cases}
0&, \text{if } \h_i\geq 0,\\
-\frac{\tau\sqrt{n}}{2\gm}\h_i^2&, \text{if } -\frac{2\gm}{\tau\sqrt{n}}\leq \h_i < 0, \\
2\frac{\gm}{\tau\sqrt{n}} + 2 \h_i&, \text{if } \h_i\leq -\frac{2\gm}{\tau\sqrt{n}}.
\end{cases}\nn
\end{align}

\noindent\textbf{Convergence of the (AO)}.
Now that the (AO) is simplified as in \eqref{eq:AOw3}, we can get a handle on the limiting behavior of the optimization itself as well as of the optimal $\tilde\w$. But first, we need to properly normalize the (AO) by dividing the objective in \eqref{eq:AOw3} by $\sqrt{n}$. Also, for convenience, redefine $\tau:=\frac{\tau}{\sqrt{\delta}}$. By the WLLN, we have
 $\frac{\gm}{\sqrt{n}}\rP \sqrt\delta$, and, for all $\tau>0$, $\frac{1}{n} \sum_{i=1}^n \upsilon(\tau;\h_i,\gm)\rP
Y(\tau) := -\frac{\tau}{2}\int_{0}^{\frac{2}{\tau}}h^2p(h)\mathrm{d}h + \frac{2}{\tau}Q\left(\frac{2}{\tau}\right) + 2\int_{\frac{2}{\tau}}^{\infty}hp(h)\mathrm{d}h.
$
With these we can evaluate the point-wise (in $\tau$) limit of the objective function in \eqref{eq:AOw3}. Next, we use the fact that the objective is convex in $\tau$ and Lemma \uniform, to conclude that the convergence is indeed uniform in $\tau$. Hence, the random optimization in \eqref{eq:AOw3} converges to the following deterministic optimization $\min_{\tau>0}\frac{\tau\delta}{2}+\frac{\sigg}{2\tau}+Y(\tau)$; some algebra shows that the latter is the same as \eqref{eq:DO_thm}.
%
If $\delta>1/2$, then, it can be shown via differentiation that the objective function of it is strictly convex. Also, it is nonnegative; thus, the entire expression in \eqref{eq:AOw3}, which is nothing but the (AO) problem we started with, converges in probability to \eqref{eq:DO_thm}.
What is more, using \consist it can be shown that the optimal $\tau_*(\g,\h)$ of the (AO) converges in probability to the unique optimal solution $\tau_*$ of \eqref{eq:DO_thm}. This is crucial for the final step of the proof.

\vp\noindent\textbf{Proving $\tilde\w\in\Sc$}.
Recall the definition in \eqref{eq:Sc}. We prove that
$\frac{1}{n}\sum_{i=1}^n\ind{\tilde\w_i\leq -1} \rP Q(1/\tau_*)$. From \eqref{eq:w_opt}, $\ind{\tilde\w_i\leq -1} = \ind{\h_i\leq -\frac{\|\g\|}{\sqrt{n}\sqrt{\delta}\tau}}$. Recall, $\|\g\|/\sqrt{n}\rP\sqrt{\delta}$ and $\tau\rP\tau_*$. Conditioning on those high probability events it can be shown that
$
\frac{1}{n}\sum_{i=1}^n\ind{\h_i\leq -\frac{\|\g\|}{\sqrt{n}\sqrt{\delta}\tau}} \rP \frac{1}{n}\sum_{i=1}^n\ind{\h_i\leq -\frac{1}{\tau_*}} \rP Q(\frac{1}{\tau_*}).
$

\vspace{-15pt}
\section{Discussion and Conclusion}\label{sec:conc}
\vspace{-8pt}
In this paper we have used the CGMT framework of \citep{COLT} to precisely compute the $\Pe$ of the box relaxation method (BRO) to recover BPSK signals in MIMO systems. At high SNR we obtain $\Pe=Q(\sqrt{(\delta-1/2)SNR})$, compared to the Mathced filter bound (MFB), $Q(\sqrt{\delta SNR})$. As the interested reader may observe and expect, similar results can be achieved for higher order constellations (m-PAM, m-QAM, m-PSK, etc.). However, we shall leave the detailed calculations for another occasion. 

In the proof outline, we made use of the set 
\begin{equation}\nn
\Sc=\{\vb:~\Big|\frac{1}{n}\sum_{i=1}^n\mathds{1}_{(\vb_i\leq -1)}-Q(\frac{1}{\tau^*})\Big|<\epsilon\},
\end{equation}
to establish that the (AO) and (PO) have the same expected BER. A study of our analysis of the (AO) reveals that error events for each of the bits in the (AO) are \emph{iid}. This means that if, for constant $k$, we define the set:
\begin{equation}\nn
\Sc_k^*=\{\vb:~\Big|\frac{1}{{{n}\choose{k}}}\sum_{\substack{T\in\{1,...,n\}\\ |T|=k}}\mathds{1}_{(\vb_{i_1}\leq -1,...,\vb_{i_k}\leq -1)}-Q^k(\frac{1}{\tau^*})\Big|<\epsilon\}
\end{equation}
then $\lim_{n\rightarrow\infty}\Pro\{\w_\phi\in S_k^*\}=1$. By Thm.~\ref{thm:main}, this implies $\lim_{n\rightarrow\infty}\Pro\{\w_\Phi\in S_k^*\}=1$, which means that error events for any fixed $k$ bits in the (PO) are also iid. This fact has significant consequences. For example, it implies that,  when a block of data is in error, only a few of its bits are. This means that the output of the \BRO can be used by various local methods to further reduce the BER. We shall explain this in future work.

\newpage
\bibliography{compbib}

\end{document}